\documentclass[3p,times,procedia]{elsarticle}
\usepackage{nupha_ecrc}

\volume{00}

\firstpage{1}

\journalname{Nuclear Physics A}

\runauth{}

\jid{nupha}

\jnltitlelogo{Nuclear Physics A}

\usepackage{amssymb}
\usepackage[figuresright]{rotating}
\usepackage{calc}
\usepackage{rotating}
\usepackage{float}
\usepackage{graphics}
\usepackage{graphicx}

\usepackage[hang,small,bf]{caption}
\usepackage{amsmath, amsthm, amssymb}
\usepackage{slashed}
\usepackage{amsfonts}
\usepackage{dsfont}

\newcommand{\psibar}{\ensuremath{\bar\psi}}

\newcommand{\pbp}{\ensuremath{\psibar\psi}}

\newcommand{\Ns}{\ensuremath{N_\sigma}}
\newcommand{\vev}[1]{\ensuremath{\left<#1\right>}}

\newcommand{\ie}{i.~e.\ }
\newcommand{\eg}{e.~g.\ }

\usepackage{multirow}
\begin{document}

\begin{frontmatter}
%% Instructions from Editor: Please use the following \dochead only in the preprint version (e-print arXiv etc.); 
%% use empty \dochead{} when submitting to Nuclear Physics A!
%\dochead{XXVIth International Conference on Ultrarelativistic Nucleus-Nucleus Collisions\\ (Quark Matter 2017)}
%% e.g. \dochead{17th International Conference on Dynamical Processes in Excited States of Solids}
\title{Topology (and axion's properties) from lattice QCD with a dynamical charm}
\author[berlin]{Florian~Burger}
\author[pappo]{Ernst-Michael Ilgenfritz}
\author[LNF]{Maria~Paola~Lombardo}
\author[berlin]{Michael~M\"uller-Preussker\textdagger}
\author[pappo]{Anton Trunin}
\address[berlin]{Physics Department, Humboldt University Berlin, Newtonstr. 15, 12489 Berlin, Germany}
\address[LNF]{Frascati National Laboratory, National Institute for Nuclear Physics, 
Via Enrico Fermi 40, 00044 Frascati (Rome), Italy}
\address[pappo]{Bogoliubov Laboratory of Theoretical Physics, Joint Institute for Nuclear Research, 
Joliot-Curie Str. 6, 141980 Dubna, Russia}
\begin{abstract}
We present results on QCD with four dynamical
flavors in the temperature range $0.9 \lesssim T/T_c \lesssim 2$. 
We have performed lattice simulations with 
Wilson fermions at maximal twist and measured the topological
charge with gluonic and fermionic methods. The topological charge
distribution is studied by means of its cumulants, which encode
relevant properties of the QCD axion, a plausible Dark Matter candidate.
The topological susceptibility measured
with the fermionic method exhibits a power-law decay 
for $T/T_c \gtrsim 2$, with an exponent close to the one predicted
by the Dilute Instanton Gas Approximation (DIGA).
Close to $T_c$ the temperature dependent 
effective exponent approaches the 
DIGA result from above, in agreement with recent 
analytic calculations. These results constrain the axion window, 
once an assumption on the fraction of axions 
contributing to Dark Matter is made.
\end{abstract}
\end{frontmatter}
% \linenumbers

\section{Topology in QCD - a Long Standing Focus of Strong Interactions}
\label{}
There are at least two compelling reasons for studying topology in QCD:
first, at high temperature  
it helps to elucidate features of the strongly interacting 
Quark Gluon Plasma~\cite{Shuryak}; 
second, it is linked with the strong CP problem, 
and in turn with its solution: the existence of axions, on which we will
comment at the end of this note. 

It is widely agreed that non-perturbative properties such as topological ones
call for lattice studies,
and that topological studies are hampered by technical difficulties 
on a discrete lattice~\cite{Muller-Preussker:2015daa}.  
However, recent methodological progress, 
together with adequate computer resources, have to some extent reopened the field,
leading to  the first results on topology at high temperature with dynamical
fermions~\cite{Trunin:2015yda,Bonati:2015vqz,Borsanyi,Petreczky:2016vrs}.  
Moreover, and very importantly, a seminal paper (still limited to the
quenched approximation) did show how to use results on 
topology and Equation of State
to constrain the mass of the post-inflactionary QCD 
axion, paving the way to more quantitative 
results~\cite{Berkowitz:2015aua}.

\section{Simulations and Observables}
We performed simulations with four flavors of maximally 
twisted mass Wilson fermions in the isospin limit, \ie with 
degenerate up and down 
quark masses, and physical strange and charm masses.  
We used three different lattice spacings 
in our finite temperature simulations:
for each lattice spacing we explored temperatures ranging 
from $150$ MeV to $500$ MeV 
by varying the temporal size of the lattice $N_t$. So far 
we have generated finite temperature configurations for eight sets of 
parameters that correspond to four values of the charged pion mass of 
about $470$, $370$, $260$ and $210$ MeV for which two, three, two and 
one value(s) of the lattice spacing (ranging from $0.0646$ fm till 
$0.0936$ fm) have been considered, respectively. 
The advantage of this approach is that we are allowed to rely 
on the setup of $T=0$ simulations of ETMC. An obvious disadvantage is
the mismatch of temperatures realised with different lattice spacings.
In the chiral sector the main quantitity of interest for this note will 
be the disconnected susceptibility of the chiral condensate, 
$\chi_{disc} = 
\Ns^3/T \left( \vev{(\bar \psi \psi)^2} - \vev{\pbp}^2 \right)$ 
shown in Fig.1 for two values of the pion mass.
\begin{figure}
\vskip -0.5 truecm
\begin{center}
\hskip -1.5truecm\includegraphics[width=8.0truecm]{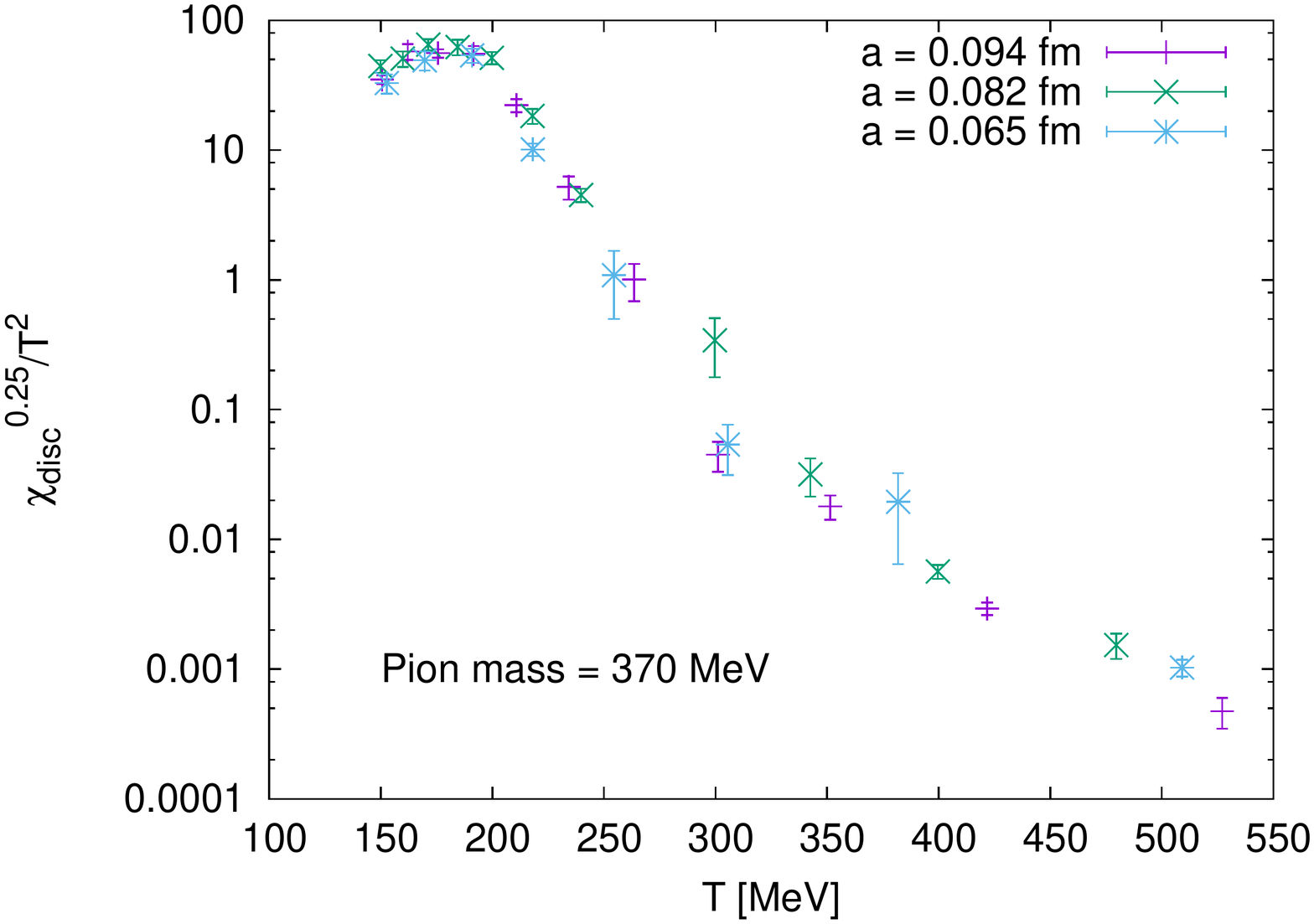}
\hskip -1.2truecm\includegraphics[width=8.0truecm]{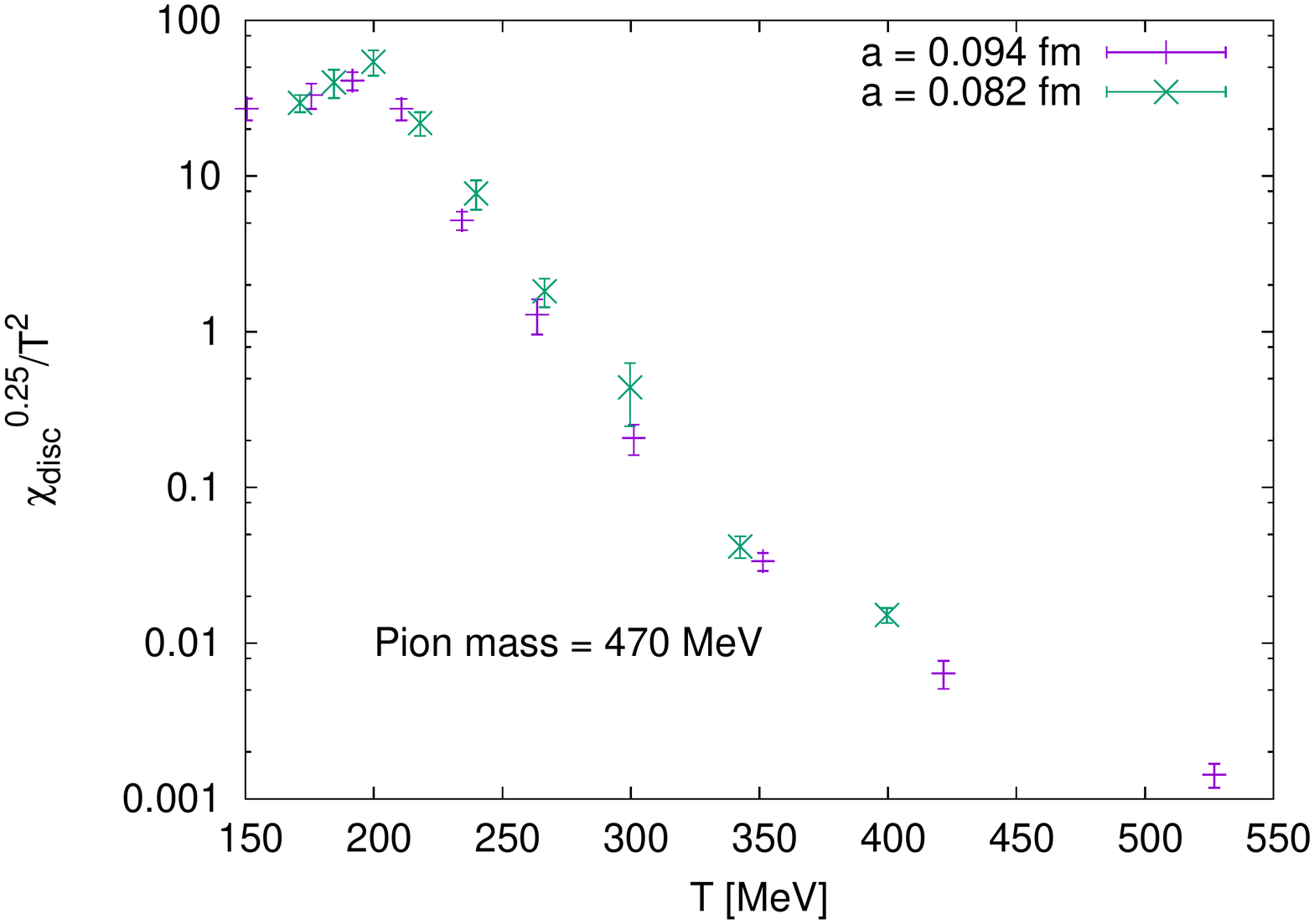}
\end{center}
\vskip -0.5 truecm
\caption{The chiral disconnected susceptibility, evaluated for two different pion masses 
and on lattices of different coarseness. The residual lattice spacing dependence is below 
the statistical errors.}
\end{figure}

\section{$\theta$ Term, Strong CP Problem and Topology}
It is well known that the QCD Lagrangian admits a CP violating term
%\begin{equation}
$ {\cal{L}}_{QCD}=  {\cal{L}}_{QCD} + \theta\frac{g^2}{32 \pi^2}  F^a_{\mu \nu} \tilde {F}^{a \rho \sigma}$
%\end{equation}
where $\frac{g^2}{32 \pi^2} F^a_{\mu \nu} \tilde{F}^{a \rho \sigma}$ 
is the topological charge density $Q(x)$. The $\theta$ term gives an electric dipole moment
to the neutron, which is strongly constrained experimentally, leading to the bound
$\theta < 10^{-10}$. The strong CP problem consists in explaining this unnaturally small value,
and we will briefly discuss the axion solution at the end. 

We present results for the second cumulant of the topological charge distribution, \ie the 
topological susceptibility  
%\begin{equation}
$~~~~\chi(T) =  -\frac{1}{V}\frac{\partial^2 \ln {\cal Z}_{QCD}(\theta,T)}{\partial \theta^2}\bigg\rvert_{\theta = 0} \equiv \frac{1}{V}  \vev{Q^2} ~~~~$
%\end{equation}   
and for the ratio of the fourth to second cumulant \\
$b_2 = -\frac{1}{12}\frac{\vev{Q^4}-3\vev{Q^2}^2}{\vev{Q^2}}$.
\begin{figure}
\hskip -1 truecm\includegraphics[width=4.5truecm,height=6.5truecm]{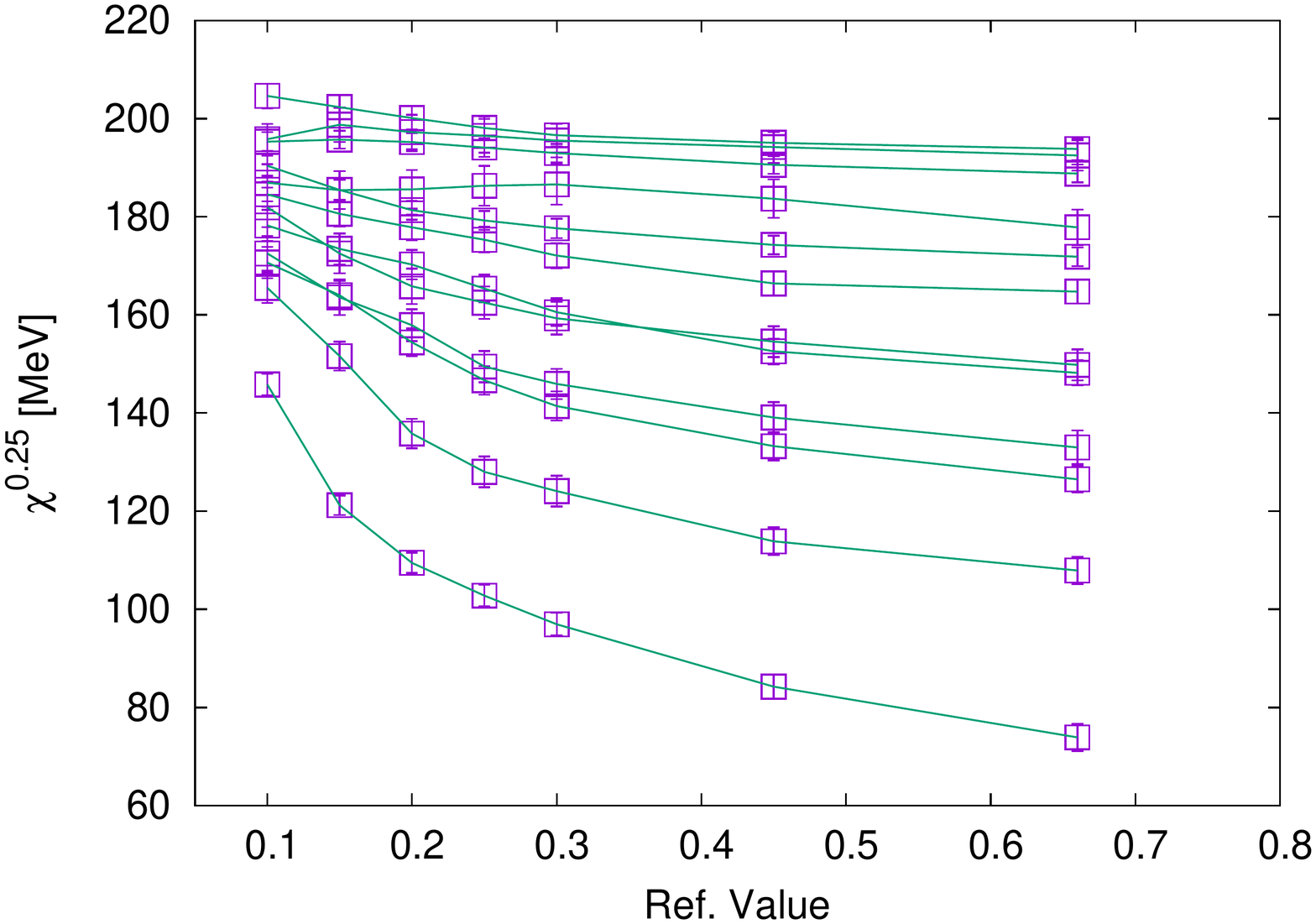}
\includegraphics[width=4.5truecm,height=6.5truecm]{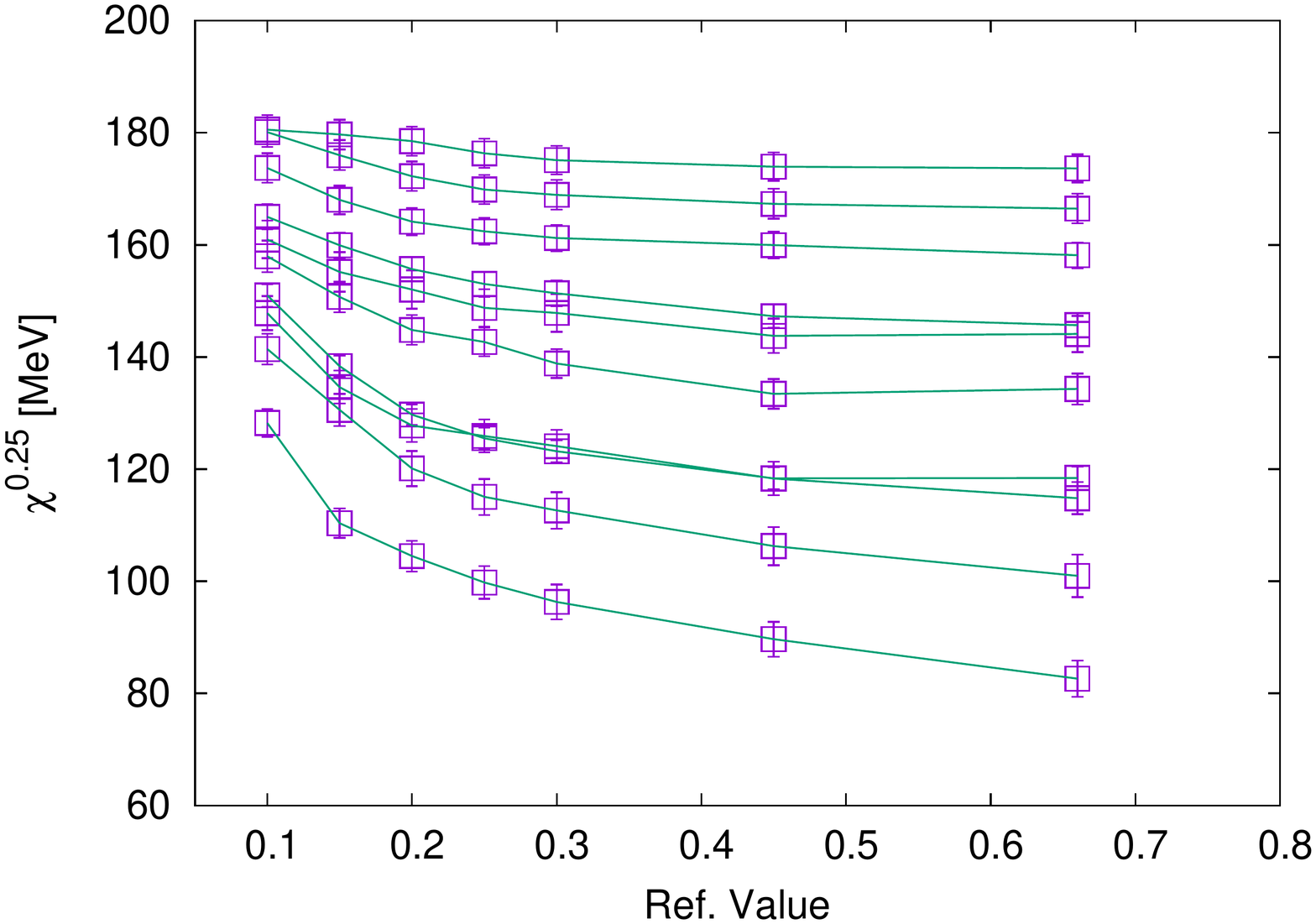}
\includegraphics[width=7truecm,height=6.5truecm]{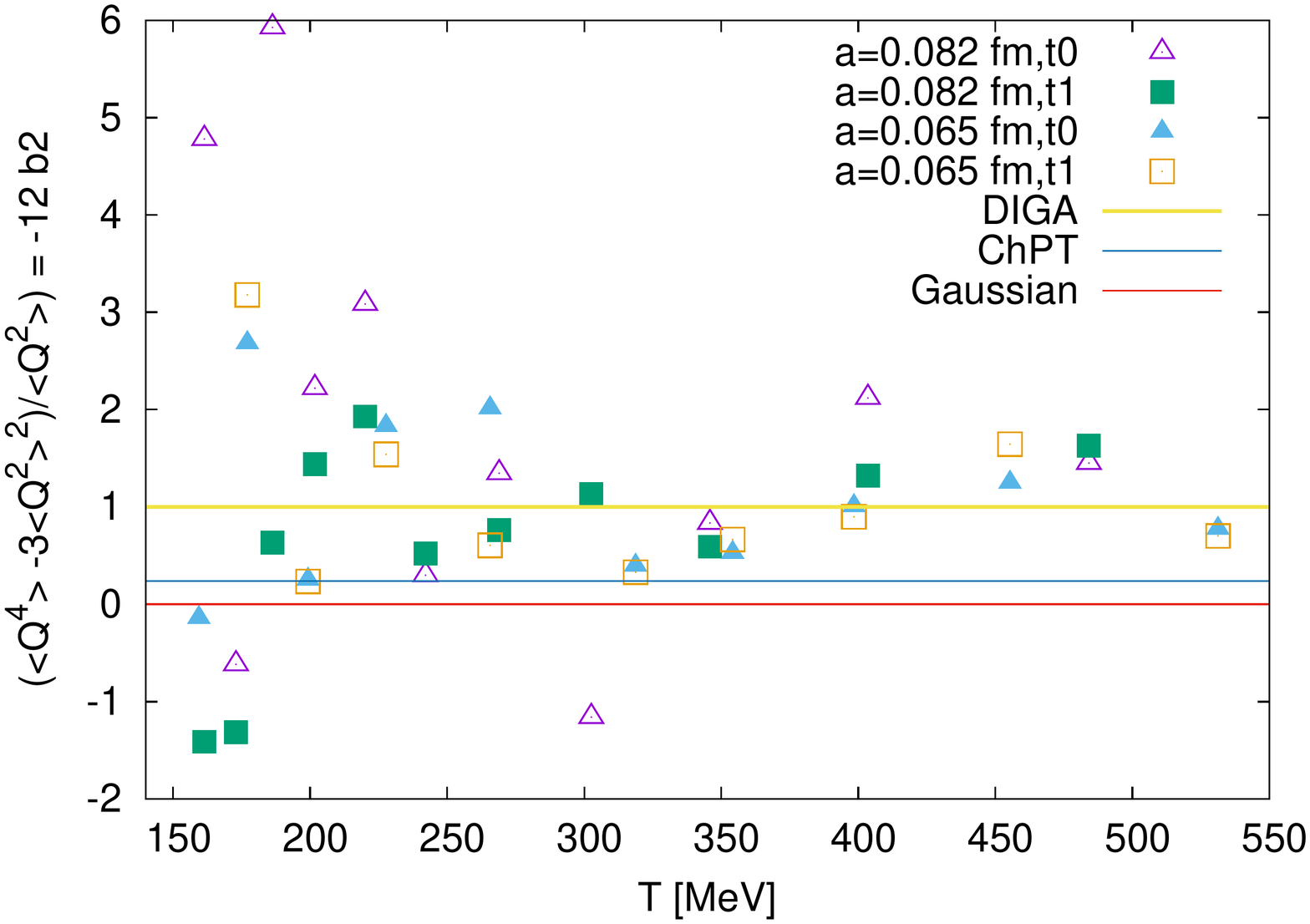}
\vskip -0.5truecm
\caption{Fourth root of the topological susceptibility as a function of the chosen 
Wilson flow reference value, for $a = 0.082$ fm ($a = 0.065$ fm) and temperatures 
in the range $[150 \ldots 500]$ MeV  decreasing from top to bottom in the left (center)
panel.
In the right panel we show  $b_2$ as a function of temperature for $t0$ and $t1$, see text.}
\end{figure}
A first set of results for the topological susceptibility  
have appeared in \cite{Trunin:2015yda}, where we relied on
the gradient flow to regulate the short distance fluctuations. 
On selected (decorrelated) configurations taken from the HMC evolution, we evolve the link 
variables in a ficticious flow time $t$, monitoring one chosen observable, in our case the 
gauge action $E$, as a function of $t$ and stopping the evolution once 
${{t_k}^2 < E>}_{t_k} = k$, with $k$ an arbitrarily chosen value. It turns out
that observables measured on the evolved links are renormalised at the scale 
$1/\sqrt{(8 t_k)}$, hence they have a well defined continuum limit.  
The procedure is demonstrated in Fig. 2 (left and center), for two different lattice spacings,
and most of our analysis uses $t_{0.3}$ and $t_{0.66}$ which, to keep the notation simple, we 
dub $t0$ and $t1$.
In Fig.2 (right panel), we present the results for the quantity $b_2$ obtained in
correpondence of $t0$ and $t1$, and for two different lattice spacings. In the same diagram 
we indicate the results from the Dilute Instanton Gas Approximation (DIGA), which should be 
reached at high temperatures, the results from Chiral Perturbation Theory (valid at low 
temperatures), and -- as a reference -- the Gaussian result. The results are dominated by 
fluctuations at low temperatures, while at high temperatures a clear pattern emerges, 
broadly consistent with the DIGA prediction.
The results for the topological susceptibility from the gluonic definition
have been presented in Ref. \cite{Trunin:2015yda}: a salient feature there is a very slow decay, 
well approximated by a linear behaviour at large temperatures.  
Further results with different quark masses show no significant quark mass dependence. 
This is calling for more detailed analysis.

\begin{figure}
\includegraphics[width=8truecm]{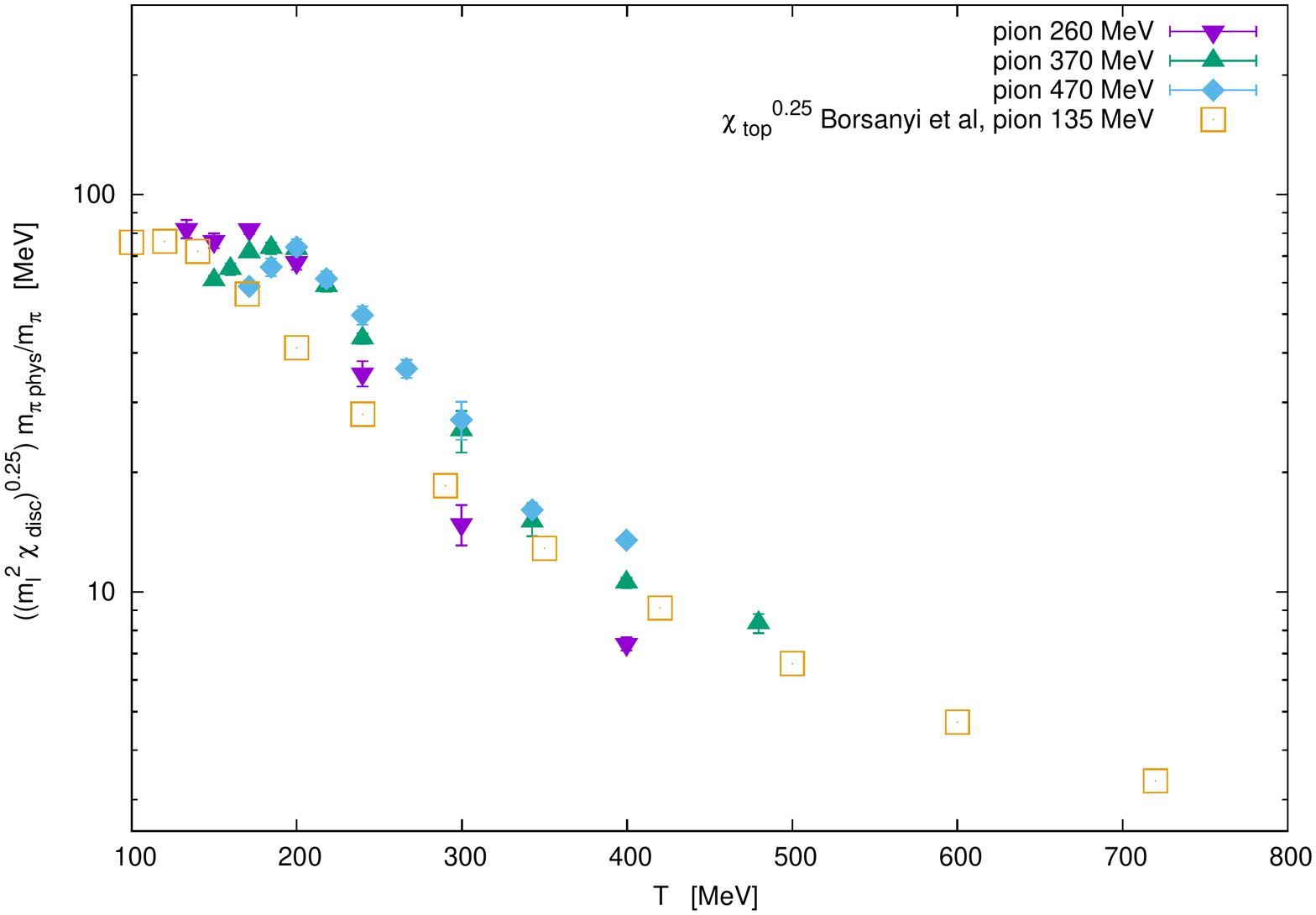}\includegraphics[width=8 truecm]{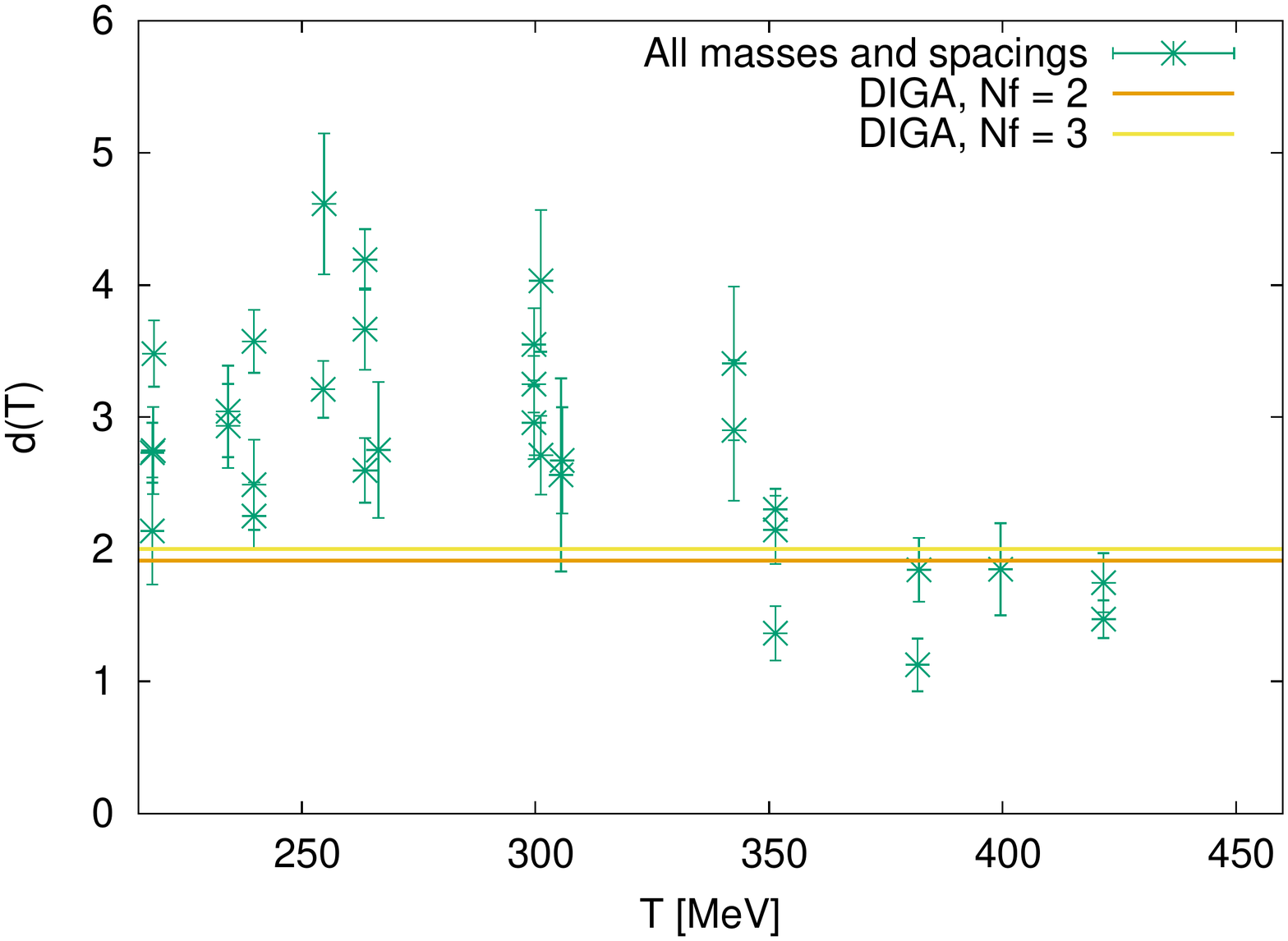} 
\vskip -0.5 truecm
\caption{The fourth root of $m_l^2 \chi_{disc}$ versus temperature for three different pion masses, 
rescaled to the physical pion mass. As discussed in the text, this quantity gives the topological 
susceptibility in the Quark Gluon Plasma phase (left).
The effective exponent describing the (local) power law behaviour of the topological susceptibility 
(right).}
\label{fig:effexp}
\end{figure}

An alternative way to measure the topological susceptibility was put forward
in Ref.~\cite{Bazavov:2012qja}: the authors note that on a smooth gauge configuration, 
if lattice artifacts are small, the topological charge and the integrated pseudo-scalar 
bilinear are related for small quark masses, hence 
$\chi_{top} = m_l^2 \chi_{5,disc}$. Previous results show that residual
$U_A(1)$ breaking above $T_c$ is small, allowing the identification
$\chi_{5,disc} = \chi_{disc}$, hence $\chi_{top} = m_l^2 \chi_{disc}$ in the Quark Gluon Plasma 
phase. Incidentally, this automatically gives a leading pion mass dependence 
$\chi_{top} \propto m_\pi^4$, in agreement with DIGA. In Figure \ref{fig:effexp} (left panel)  
we present the results for the topological susceptibility obtained in this way. 
Finally, we explore in some more detail the functional dependence 
of the topological susceptibility on the
temperature. As models predict a power-law fall off $\chi(T) \propto T^{-d}$, we search 
for this by defining a local effective power \cite{Aarts:2010ek} 
$d_{eff}(T) = T d/dT \log \chi(T)$ which we show in Fig. \ref{fig:effexp} (right panel).
It approaches the DIGA results above $T \sim 350$ MeV and has a larger value for smaller 
temperature, in agreement with the proposal of Ref.~\cite{Larsen:2017rjs}. It is worth 
noticing that a larger exponent close to $T_c$ follows naturally from the Chiral Equation 
of State~\cite{Ejiri:2009ac}.

\section{A Window on the Axion}
An elegant solution to the strong CP problem  postulates an extra particle, 
a pseudo-Goldstone boson of the spontaneously broken Peccei-Quinn symmetry, 
which couples to the QCD topological charge, with a coupling suppressed by 
a scale $f_a$. At leading order in $1/f_a$ the axion can be treated as an
external source, and the mass is given by
$m_a^2(T) f_a^2 = \frac{\partial^2 F(\theta, T)}{\partial \theta^2} \equiv
\chi(T)$. 
As the cumulants of the topological charge distribution are related (via the 
Edgeworth expansion) to the Taylor coefficients of the expansion of the free energy
around $\theta=0$, higher order cumulants and their ratio such as $b_2$ carry information 
on the axion's interactions.

Axions  make ideal Dark Matter candidates, and experiments are actively searching for them. 
Experiments need an acceptable mass range to explore, which is provided by a combination 
of QCD results, phenomenology and experimental constraints:  
as the Universe is cooling down, both the Hubble parameter and the topological
susceptibility, and hence the axion mass, are changing.
When the axion mass is of the order of the inverse of the Hubble parameter, it ``freezes out'',  
$3 H(T) = m_a(T) = \sqrt{\chi(T)}/f_a$. The admitted range of coupling $f_a$ is given by  
phenomenology, in particular taking into account that the density of Dark Matter
 should be not less that the corresponding axion density (overclosure bound), 
see \eg Ref.~\cite{Villadoro}. Our results using the fermionic definition
of $\chi_{t}$
are at the moment consistent with an absolute lower bound for the axion mass of about 
25 $\mu$eV for a saturated overclosure bound. 
A more realistic estimate taking into account further contributions to Dark Matter 
pushes this lower bound to larger values, into a region not yet accessible to experiments, 
while slower decays (as those observed by us using the gluonic definition) would lower the 
axion bound \cite{Villadoro}. Our results using the fermionic operator reach the 
DIGA exponent from above,  and are broadly consistent with those of 
Refs.~\cite{Borsanyi,Petreczky:2016vrs} for $T > 300$ MeV, once rescaled to the physical pion mass.
However, the reasons behind the discrepancies with our own gluonic results, 
Ref.~\cite{Trunin:2015yda}, and with those of Ref.~\cite{Bonati:2015vqz}  
are not fully understood, and in our opinion the present results cannot
yet be considered conclusive. 

\vskip 0.1 truecm
We are grateful to the Supercomputing Center of Lomonosov Moscow State
University, to the HybriLIT group of JINR and to CINECA 
(INFN-CINECA agreement) for computational resources. E.-M.~I. and M.P.~L.
wish to thank the Director and the Staff at the European Center for Theoretical 
Nuclear Physics, Trento, for their kind support and hospitality.


\begin{thebibliography}{00}
\bibitem{Shuryak}
E.~Shuryak,
%``Physics of Strongly coupled Quark-Gluon Plasma,''
Prog.\ Part.\ Nucl.\ Phys.\  {\bf 62} (2009) 48
\bibitem{Muller-Preussker:2015daa}
M.~M\"uller-Preussker, PoS LATTICE {\bf 2014}, 003 (2015)
\bibitem{Trunin:2015yda}
A.~Trunin, F.~Burger, E.~M.~Ilgenfritz, M.~P.~Lombardo and M.~M\"uller-Preussker,
%``Topological susceptibility from $N_f=2+1+1$ lattice QCD at nonzero temperature,''
J.\ Phys.\ Conf.\ Ser.\  {\bf 668} (2016) no.1,  012123
\bibitem{Bonati:2015vqz}
C.~Bonati, M.~D'Elia, M.~Mariti, G.~Martinelli, M.~Mesiti, F.~Negro, F.~Sanfilippo and G.~Villadoro,
%``Axion phenomenology and $\theta$-dependence from $N_f = 2+1$ lattice QCD,''
JHEP {\bf 1603} (2016) 155
\bibitem{Borsanyi}
S.~Borsanyi {\it et al.},
%``Calculation of the axion mass based on high-temperature lattice quantum chromodynamics,''
Nature {\bf 539} (2016) no.7627, 69
\bibitem{Petreczky:2016vrs}
P.~Petreczky, H.~P.~Schadler and S.~Sharma,
%``The topological susceptibility in finite temperature QCD and axion cosmology,''
Phys.\ Lett.\ B {\bf 762} (2016) 498
\bibitem{Berkowitz:2015aua}
E.~Berkowitz, M.~I.~Buchoff and E.~Rinaldi,
%``Lattice QCD input for axion cosmology,''
Phys.\ Rev.\ D {\bf 92} (2015) no.3,  034507
\bibitem{Bazavov:2012qja}
A.~Bazavov {\it et al.} [HotQCD Collaboration],
%``The chiral transition and $U(1)_A$ symmetry restoration from lattice QCD using Domain Wall Fermions,''
Phys.\ Rev.\ D {\bf 86} (2012) 094503
\bibitem{Aarts:2010ek}
G.~Aarts, S.~Kim, M.~P.~Lombardo, M.~B.~Oktay, S.~M.~Ryan, D.~K.~Sinclair and J.-I.~Skullerud,
%``Bottomonium above deconfinement in lattice nonrelativistic QCD,''
Phys.\ Rev.\ Lett.\  {\bf 106} (2011) 061602
\bibitem{Larsen:2017rjs}
R.~Larsen and E.~Shuryak,
%``Correlations and fluctuations of the gauge topology at finite temperatures,''
arXiv:1703.02434 [hep-ph]
\bibitem{Ejiri:2009ac}
S.~Ejiri et al.,
%``On the magnetic equation of state in (2+1)-flavor QCD,''
Phys.\ Rev.\ D {\bf 80} (2009) 094505
\bibitem{Villadoro}
G.~Grilli di Cortona, E.~Hardy, J.~Pardo Vega and G.~Villadoro,
%``The QCD axion, precisely,''
JHEP {\bf 1601} (2016) 034.
\end{thebibliography}
\end{document}